\documentclass[twocolumn]{aastex701}

\begin{document}

\title{Discovery of CO Clouds Associated with the X-ray Jets of SS 433: Evidence for Shock--Cloud Interaction Enhancing Nonthermal X-ray Emission}

\author[0000-0002-4037-1346]{Haruka Sakemi}
\affiliation{Graduate School of Sciences and 
Technology for Innovation, Yamaguchi University, 1677-1 Yoshida, Yamaguchi, 753-0841, Japan}
\affiliation{Nobeyama Radio Observatory, National Astronomical Observatory of Japan (NAOJ), National Institutes of Natural Sciences
(NINS), 462-2, Nobeyama, Minamimaki, Minamisaku, Nagano 384-1305, Japan}
\email{sakemi@yamaguchi-u.ac.jp}

\author[0000-0003-2062-5692]{Hidetoshi Sano} 
\affiliation{Faculty of Engineering, Gifu University, 1-1 Yanagido, Gifu 501-1193, Japan}
\email{sano.hidetoshi.w4@f.gifu-u.ac.jp}

\author[0000-0002-8966-9856]{Yasuo Fukui} 
\affiliation{Department of Physics, Nagoya University, Furo-cho, Chikusa-ku, Nagoya 464-8601, Japan}
\email{fukui@a.phys.nagoya-u.ac.jp}

\author[0000-0001-6353-7639]{Mami Machida}
\affiliation{Division of Science, National Astronomical Observatory of Japan, 2-21-1 Osawa, Mitaka, Tokyo 181-8588, Japan}
\email{mami.machida@nao.ac.jp}

\author[0000-0003-2579-7266]{Shigeo S. Kimura}
\affiliation{Frontier Research Institute for Interdisciplinary Sciences, Tohoku University, Sendai 980-8578, Japan}
\affiliation{Astronomical Institute, Graduate School of Science, Tohoku University, Sendai 980-8578, Japan}
\email{shigeo@astr.tohoku.ac.jp}

\author[0000-0003-3990-1204]{Masato I.N. Kobayashi}
\affiliation{Meta-Hierarchy Dynamics Unit, National Institute for Fusion Science, 322-6 Oroshi-cho, Toki, Gifu 509-5292. Japan}
\email{kobayashi.masato@nifs.ac.jp}

\author[0000-0002-3562-6965]{Kazuho Kayama}
\affiliation{Department of Physics, Kyoto University, Kitashirakawa Oiwake-cho, Sakyo, Kyoto 606-8502, Japan}
\affiliation{Remote Sensing Technology Center of Japan, 3-17-1, Toranomon, Minato-ku, Tokyo, 105-0001, Japan}
\email{kayama.kazuho.57r@kyoto-u.jp}

\author[0000-0001-5792-3074]{Hiroaki Yamamoto}
\affiliation{Graduate School of Science, Nagoya University, Furo-cho, Chikusa-ku, Nagoya, Aichi 464-8602, Japan}
\email{hiro@a.phys.nagoya-u.ac.jp}

\author[0000-0002-1411-5410]{Kengo Tachihara}
\affiliation{Graduate School of Science, Nagoya University, Furo-cho, Chikusa-ku, Nagoya, Aichi 464-8602, Japan}
\email{k.tachihara@a.phys.nagoya-u.ac.jp}

\author[0000-0003-0292-3645]{Hiroshi Nagai}
\affiliation{National Astronomical Observatory of Japan, 2-21-1 Osawa, Mitaka, Tokyo 181-8588, Japan}
\affiliation{The Graduate University for Advanced Studies, SOKENDAI, 2-21-1 Osawa, Mitaka, Tokyo}
\email{hiroshi.nagai@nao.ac.jp}

\begin{abstract}
We report the first identification of molecular clumps directly associated with the re-brightening regions of the large-scale X-ray jets of SS 433, based on $^{12}$CO ($J$ = 1--0) observations with the Nobeyama 45-m Radio Telescope. Multiple clumps are detected toward the eastern and western jet heads, showing clear spatial correlation with the X-ray emission. The X-ray emission peaks immediately downstream of the molecular clumps, while the hardness ratio is enhanced at their surfaces, indicating that the observed structures cannot be explained by absorption effects. These results provide direct evidence for shock--cloud interactions between the jets and the surrounding interstellar medium. We suggest that turbulence generated at the jet--cloud interface amplifies magnetic fields, producing the observed non-thermal X-ray emission. 
Our findings highlight the importance of jet--ISM interactions in shaping the X-ray properties of microquasar jets.
\end{abstract}


\section{Introduction} \label{sec:intro}
SS 433 is one of the most active microquasars in our Galaxy, consisting of a compact object and an A-type supergiant in a binary system\citep{abell1979,margon1984,gies2002,fabrika2004,hillwig2008,kubota2010}. Jets are observed from radio to X-ray wavelengths within arcsecond scales of the source; however, the X-ray emission was previously thought to fade within $\sim$ 2 arcsec from the binary \citep{blundell2018,marti2018, migliari2002,migliari2005,sakai2025}. 

At larger distances, bright X-ray jets are detected at $\sim$ 17 arcmin east and west of SS 433. These structures, first identified with ROSAT and later observed with multiple X-ray observatories, trace relics of past jet activity \citep{brinkmann1996,safi-harb1997,safi-harb1999,moldowan2005,brinkmann2007,safi-harb2022,sunyaev2026}. The base regions of these structures, previously referred to as e0 and w1, are hereafter called the eastern and western heads. These regions provide key diagnostics of the jet history, including possible quiescent phases. Further downstream, the jets exhibit knot-like structures, referred to as the lenticular feature in the eastern jet and w1.5 and w2 in the western jet (see \citealt{kayama2025} for more details).

The head regions exhibit (i) sharply bounded X-ray morphologies and (ii) spectra dominated by non-thermal emission. Photon indices $\Gamma$ of 1.54 and 1.64 in the 1.0--7.0 keV band have been reported for the eastern and western heads, respectively \citep{kayama2022,kayama2025}, suggesting re-brightening due to recollimation shocks that accelerate high-energy electrons via diffusive shock acceleration (DSA) \citep{komissarov1997}. This interpretation is broadly consistent with recent detections of very-high-energy gamma rays by H.E.S.S. and LHAASO, including a downstream softening trend indicative of an electron origin \citep{hess2024,lhaaso2025}.

However, recollimation shocks are not the only mechanism that can account for the re-brightening of the X-ray emission. Alternatively, magnetic field amplification driven by turbulence induced by interactions between the jet and the interstellar medium (ISM) can also explain the observed X-ray emission. Such a mechanism has also been reported in supernova remnants (SNRs), where shock--cloud interactions lead to enhanced synchrotron X-ray emission \citep{fukui2003,sano2010,fukui2012,inoue2012,sano2020}. A similar process may operate in the X-ray jets of SS 433. 

Jet--ISM interactions have also been discussed in several other Galactic microquasars and X-ray binaries \citep{motta2025,arti2025,mariani2025,grollimund2026}. In particular, molecular-line studies of such systems have revealed spatial associations between dense molecular gas and radio jet relics or lobes \citep{tetarenko2018,tetarenko2020,bosch-cabot2026}. In SS 433 itself, previous studies of interstellar material have mainly focused on the outer radio shell W50, where molecular clouds possibly associated with the shell have been reported \citep{yamamoto2008,su2018,liu2020,yamamoto2022,sakemi2023,yamamoto2024,oka2026}. However, no direct association between molecular gas and the X-ray jet heads has been established. The molecular cloud N4 reported by \cite{yamamoto2008,yamamoto2022,yamamoto2024} overlaps with the western jet in projection, but it is offset from the western head region, and no corresponding molecular counterpart has been identified toward the eastern head.


In this Letter, we report $^{12}$CO ($J$ = 1--0) observations of the eastern and western head regions obtained with the Nobeyama 45-m Radio Telescope.
We detect molecular cloud clumps spatially associated with the X-ray heads, providing evidence that shock--cloud interactions, rather than recollimation shocks, may be responsible for the re-brightening of the X-ray emission.
Section \ref{sec:obs} describes the observations, Section \ref{sec:result} presents the molecular cloud properties and their spatial relationship with the X-ray jets, Section \ref{sec:discussion} discusses the implications, and Section \ref{sec:conclusion} summarizes our conclusions.

\begin{figure*}[ht]
\plotone{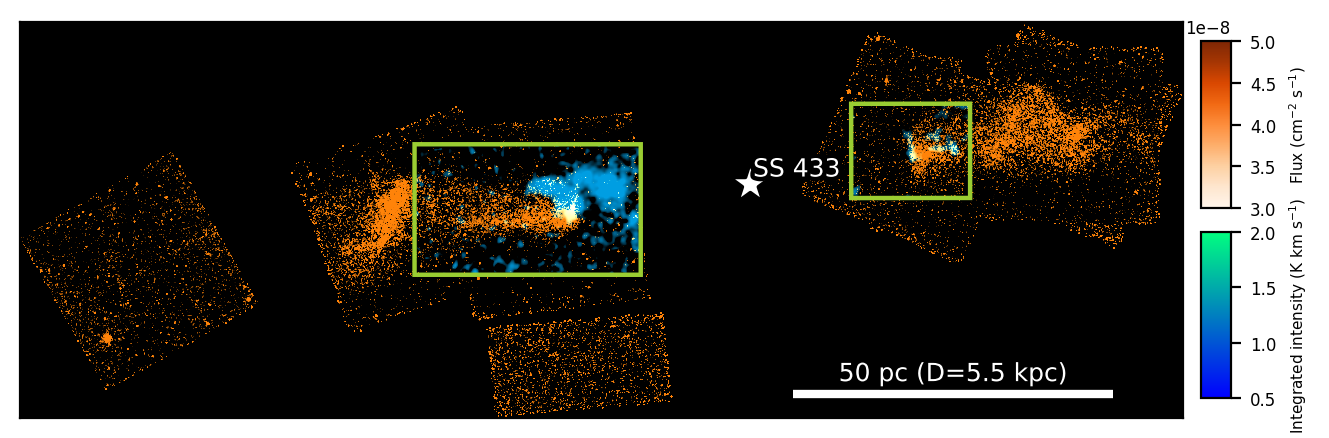}
\caption{Two-color composite image of the integrated intensity of $^{12}$CO ($J$ = 1--0) (cyan) and the X-ray photon flux in the 0.5--7 keV band observed with Chandra (orange) \citep{tsuji2025}. The velocity ranges used for the CO integration are 46.5--58.3 km s$^{-1}$ for the eastern side and 51.7--56.5 km s$^{-1}$ for the western side. The position of SS 433 is indicated by a white star. The scale bar at the lower right assumes a distance of 5.5 kpc \citep{blundell2004}. The light green boxes indicate the regions shown in Figure \ref{fig:fig2}, corresponding to the areas observed with the Nobeyama 45-m Radio Telescope in this study.}
\label{fig:fig1}
\end{figure*}

\begin{figure*}[ht]
\plotone{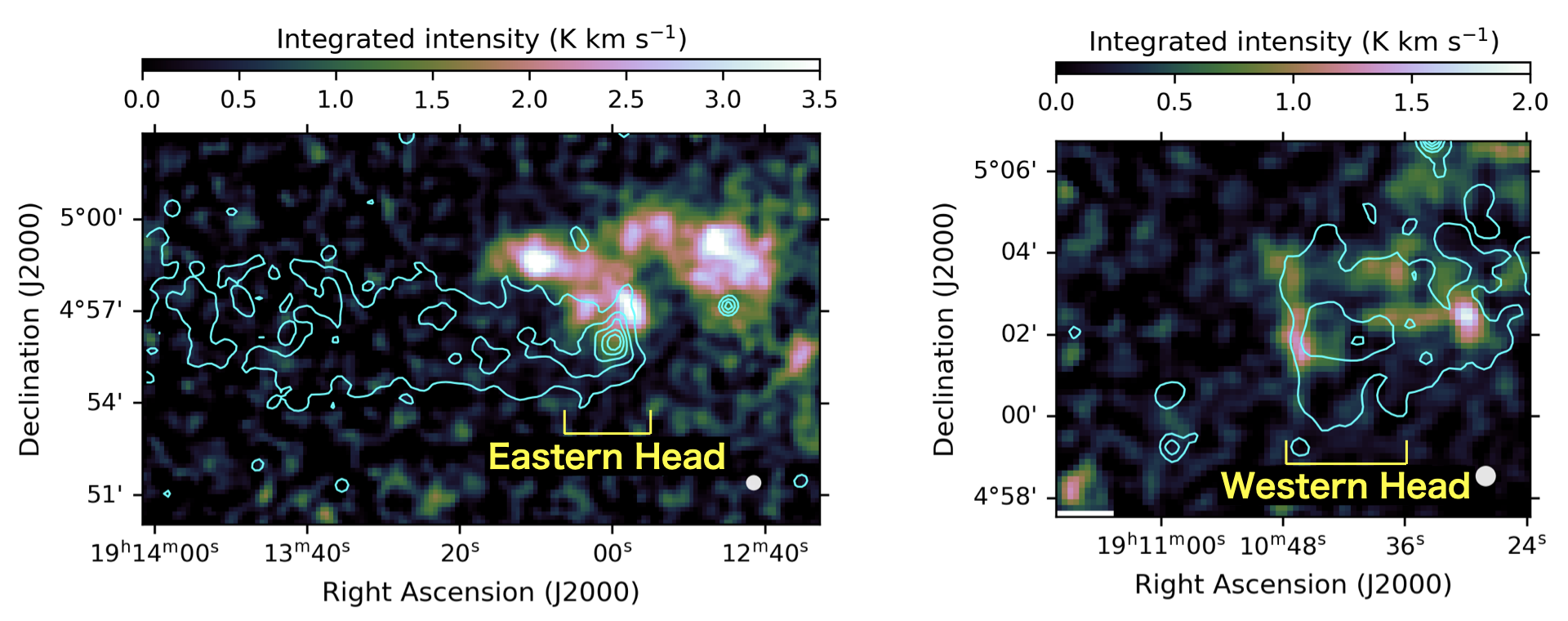}
\caption{Integrated intensity maps of $^{12}$CO ($J$ = 1--0). The left and right panels show the eastern and western jet head regions, respectively. The velocity ranges are the same as those in Figure \ref{fig:fig1}. The beam size after smoothing (30 arcsec) is shown in the lower right. Cyan contours indicate the X-ray photon flux shown in Figure \ref{fig:fig1}, with five levels ranging from 3.0 $\times$ 10$^{-8}$ to 1.2 $\times$ 10$^{-7}$ cm$^{-1}$ s$^{-1}$.}
\label{fig:fig2}
\end{figure*}

\begin{figure*}[ht!]
\includegraphics[width=1.0\linewidth]{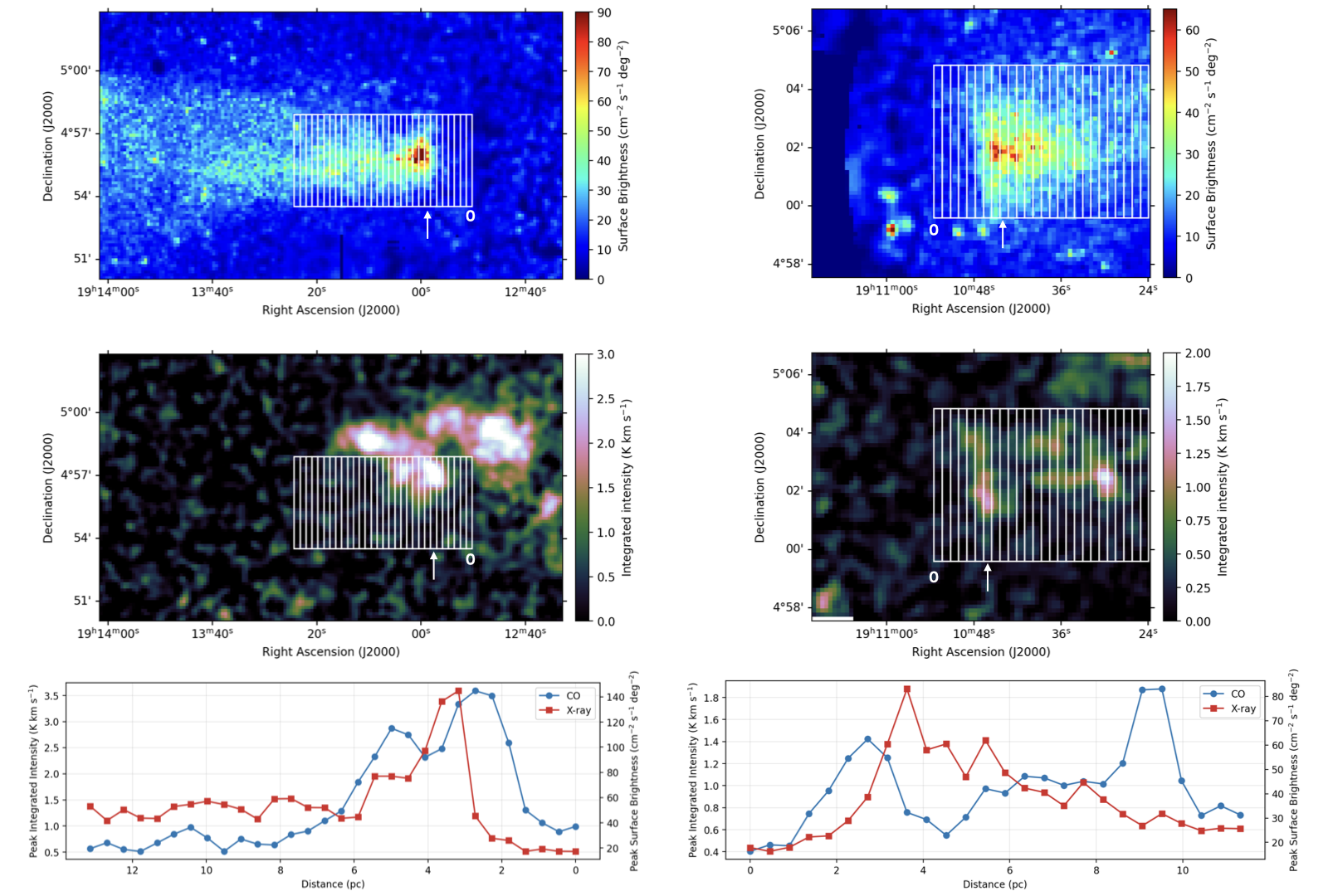}
\caption{Comparison of the spatial distributions of the X-ray surface brightness in the 0.5--10 keV band observed with XMM-Newton \citep{kayama2022,kayama2025} and the integrated intensity of $^{12}$CO ($J$ = 1--0). The X-ray images are regridded to match the spatial grid of the CO data. The left and right columns correspond to the eastern and western jets, respectively. The top panels show the X-ray surface brightness, and the middle panels show the CO integrated intensity over the same velocity ranges as in Figure \ref{fig:fig1}. The slices shown in the top and middle panels have a width of 17 arcsec, corresponding to $\sim$0.45 pc at a distance of 5.5 kpc. The white arrows indicate the slices where the first peak appears downstream from the X-ray jet head regions. The bottom panels plot the peak values of the CO integrated intensity and X-ray surface brightness within each slice. The horizontal axis represents the distance (pc) from the slice labeled 0 in the top and middle panels.}
\label{fig:fig3}
\end{figure*}

\section{Observations and Data Reduction} \label{sec:obs}
We obtained the $^{12}$CO ($J$ = 1--0) data with the 45 m telescope of the Nobeyama Radio Observatory (NRO), with observations conducted between November 2022 and December 2025.
We scanned target areas in on-the-fly mapping mode \citep{sawada2008}. 
We scanned in the right ascension and declination directions to suppress the scanning effect. 
The four-beam receiver FOREST \citep{minamidani2016} and the autocorrelation spectrometer SAM45 \citep{kuno2011} were used. 
The standard data reduction was carried out using NOSTAR.
The pointing accuracy was checked every 2 hours by observing R Aquilae ($\alpha_{J2000}$, $\delta_{J2000}$) = (19$^{\rm h}$ 06$^{\rm m}$ 22$^{\rm s}$.25, +8$^{\circ}$13$^{\prime}$48.0$^{\prime\prime}$) with the frontend H40 or Z45 \citep{{nakamura2015}}. 
All observations were conducted in equatorial coordinates. 
We used a chopper wheel to obtain the antenna temperature $T^{*}_{a}$ \citep{kutner1981}. 
We observed W51 as a standard source to fix gain variations. 
The spectral intensity was calibrated and converted to the $T_{\rm MB}$ scale by applying a main beam efficiency $\eta_{\rm MB}$ corresponding to each polarization, frequency, and observing season provided by the observatory. 
To improve the sensitivity, the spatial resolution was smoothed, resulting in a final spatial resolution of 30 arcseconds. 
The spatial and velocity grids had sizes of 8.5 arcsec and 0.2 km s$^{-1}$, respectively. 
The velocity coverage was from -45 to 113 km s$^{-1}$. 
The root-mean-square (rms) noise levels in $T_{\rm MB}$ were $\sim$ 0.18 and $\sim$ 0.15 K at 115 GHz for eastern and western regions, respectively.

\begin{figure*}[ht!]
\plotone{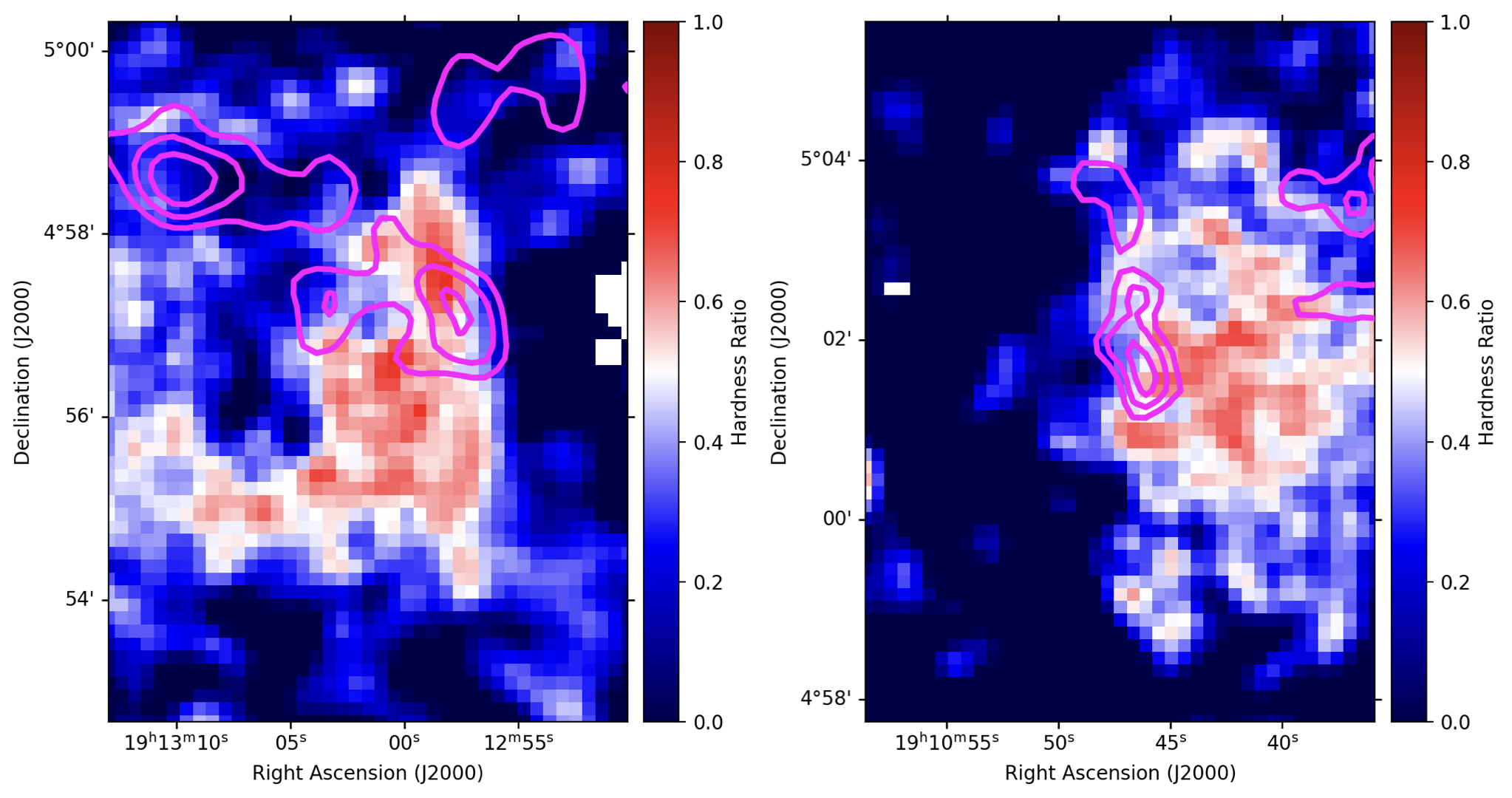}
\caption{Hardness ratio map between the 0.5--1.5 keV and 2.0--7.0 keV bands observed with XMM-Newton \citep{kayama2022,kayama2025}. The images are regridded to match the spatial grid of the 
$^{12}$CO ($J$ = 1--0) data, as in Figure \ref{fig:fig3}. Magenta contours represent the CO integrated intensity over the same velocity ranges as in Figure \ref{fig:fig1}. For the eastern side, three contour levels are shown from 2.1 to 3.5 K km s$^{-1}$, while for the western side, six levels are shown from 0.8 to 2.0 K km s$^{-1}$.}
\label{fig:fig4}
\end{figure*}

\section{Results} \label{sec:result}

Figure \ref{fig:fig1} shows the integrated intensity map of 
$^{12}$CO ($J$ = 1--0) overlaid on a two-color composite X-ray image in the 0.5--7 keV band obtained with Chandra \citep{tsuji2025}. The velocity ranges used for integration are 46.5--58.3 km s$^{-1}$ for the eastern side and 51.7--56.5 km s$^{-1}$ for the western side. Molecular clouds are clearly detected within similar velocity ranges toward the head regions of both the eastern and western jets.

Figure \ref{fig:fig2} presents zoomed-in views of the boxed regions in Figure \ref{fig:fig1}, showing the integrated intensity maps of $^{12}$CO ($J$ = 1--0) over the same velocity ranges. Multiple molecular clumps are identified in the head regions of both jets. Assuming a distance of 5.5 kpc \citep{blundell2004}, their typical sizes are $\sim$ 2 pc. In particular, the clumps located at ($\alpha_{J2000}$, $\delta_{J2000}$) = (19$^{\rm h}$ 12$^{\rm m}$ 57$^{\rm s}$.04, +4$^{\circ}$57$^{\prime}$03.6$^{\prime\prime}$) in the eastern jet and ($\alpha_{J2000}$, $\delta_{J2000}$) = (19$^{\rm h}$ 10$^{\rm m}$ 46$^{\rm s}$.11, +5$^{\circ}$01$^{\prime}$26.0$^{\prime\prime}$) in the western jet exhibit elongated morphologies along the north--south direction, aligned with the jet heads.

\begin{figure*}[ht!]
\plotone{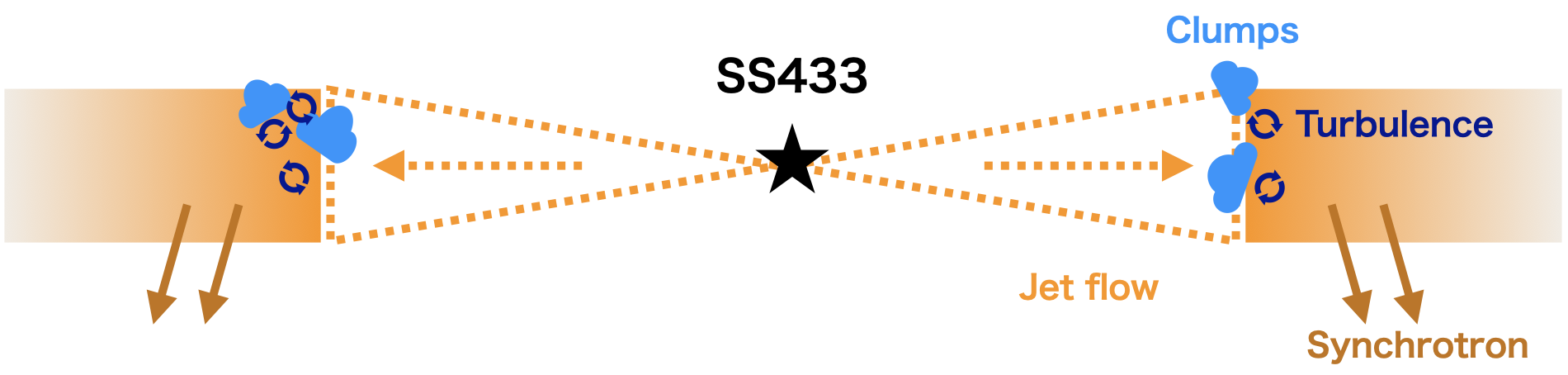}
\caption{Schematic illustration of the shock--cloud interaction occurring in the head regions of the X-ray jets of SS 433. 
}
\label{fig:fig5}
\end{figure*}

Figure \ref{fig:fig3} shows the spatial relationship between the CO emission and the X-ray jets. The left and right panels correspond to the eastern and western jets, respectively. The top panels show X-ray images in the 0.5--10 keV band observed with the XMM-Newton \citep{kayama2022,kayama2025}, while the middle panels show the integrated intensity of $^{12}$CO ($J$ = 1--0). The slices indicated in the top and middle panels have a width of 17 arcsec (corresponding to $\sim$ 0.45 pc at 5.5 kpc). The bottom panels plot the peak surface brightness of the X-ray and the peak integrated intensity of the CO emission within each slice. The horizontal axis represents the distance (pc) from the slice labeled 0 in the top and middle panels; note that the distance increases from west to east for the eastern jet, and from east to west for the western jet.

As clearly seen in the bottom panels, the X-ray emission peaks immediately downstream of the molecular clump peaks. This demonstrates a clear spatial correlation between the X-ray jets and the molecular clumps. This is the first identification of interstellar material that shows a direct spatial correlation with the X-ray jets.

Figure \ref{fig:fig4} shows the hardness ratio map between the 0.5--1.5 keV and 2.0--7.0 keV bands observed with the XMM-Newton \citep{kayama2022,kayama2025}. The hardness ratio is defined as $HR=(H-S)/(H+S)$, where $H$ and $S$ represent the surface brightnesses in the 2.0-7.0 keV and 0.5-1.5 keV bands, respectively. The contours represent the integrated intensity of $^{12}$CO ($J$ = 1--0). In the head regions of both jets, a large fraction of the area exhibits hardness ratios exceeding 0.5, indicating that hard, non-thermal X-ray emission dominates, consistent with previous studies. Notably, the peaks of the molecular clump emission are spatially offset from the regions with high hardness ratios by $\sim$ 0.5 -- 1 pc. If the high hardness ratio regions coincided with the molecular clump peaks, the enhancement could be attributed to absorption of soft X-rays by the molecular gas. However, the observed spatial offset clearly indicates a different origin. This interpretation is discussed in Section \ref{sec:discussion}.

\section{Discussion} \label{sec:discussion}
\cite{safi-harb2022} estimated that the equipartition magnetic field strength in the eastern X-ray jet head is at least $\sim$ 12 $\mu$G. They also derived a required energy of 2.4$\times$10$^{44}$ erg to accelerate electrons up to X-ray-emitting energies and a radiative cooling timescale of $\sim$ 10$^{3}$ yr. Since this timescale is much shorter than the system age ($\sim$ 10$^{4}$--10$^{5}$ yr), continuous or repeated particle re-acceleration in the head region is required. However, the photon indices observed in the head regions of the eastern and western X-ray jets, $\Gamma$ = 1.54 and 1.64, are significantly harder than those typically observed in shock regions of SNRs. In the case of SNRs, ideal DSA predicts $\Gamma$ = 1.5, whereas observed values are typically in the range of $\Gamma$ = 2--3 \citep{reynolds2008}. This discrepancy is generally attributed to synchrotron cooling and nonlinear effects operating in realistic acceleration environments. Therefore, explaining the unusually hard photon indices in the X-ray jet head regions requires invoking mechanisms that can overcome these effects.

We report the first identification of molecular clumps that are highly likely to be directly interacting with the X-ray jets of SS 433. Two key observational features are found: (i) spatial offsets between the peaks of the $^{12}$CO ($J$ = 1--0) emission and the X-ray emission, and (ii) offsets between the CO peaks and regions of high hardness ratio, with the latter preferentially enhanced at the surfaces of the molecular clumps. These characteristics can be naturally explained by a shock--cloud interaction scenario as shown in Figure \ref{fig:fig5} (e.g., \citealt{inoue2012}). A similar interpretation has been successfully applied to the SNR RX J1713.7-3946, where spatial offsets between X-ray and CO emission are observed \citep{sano2010,fukui2012,sano2020}.

In this scenario, molecular clouds with relatively high densities ($\sim$ 10$^{3}$ cm$^{-3}$) pre-exist in the head regions of both jets. As the jet propagates outward from SS 433, it collides with these clouds. The lower-density outer layers of the clouds are expected to be stripped by the jet, while the dense clumps can survive the interaction. 
In this picture, the CO emission traces the dense molecular clumps that survive the jet impact, whereas the synchrotron X-ray emission is enhanced primarily in the turbulent interaction layers around the clumps. Since magnetic-field amplification occurs at the clump surfaces or in the surrounding shear layers, rather than in the densest CO-bright cores themselves, the X-ray and CO peaks are not expected to coincide exactly. The observed offsets between the CO peaks, X-ray peaks, and high-hardness-ratio regions can therefore be naturally interpreted as a consequence of jet-driven turbulence and magnetic-field amplification around dense clumps.
This scenario suggests that the re-brightening of the X-ray jets may arise from a mechanism distinct from recollimation shocks. A similar spatial relationship has been reported in the supernova remnant RX J1713.7-3946, where pc-scale molecular clouds show offsets with respect to the X-ray structures. At sub-pc resolution, however, a more pronounced spatial anti-correlation between the X-ray emission and molecular clouds is observed\citep{sano2020}. The molecular clumps identified in the head regions of the SS 433 X-ray jets may exhibit a similar trend when observed at higher spatial resolution. 

Here, we examine whether pre-existing molecular clouds with densities of $\sim$ 10$^{3}$ cm$^{-3}$ can survive continuous interactions with the jet over timescales comparable to the jet age. Using Equation (A5) of \cite{inoue2012}, and assuming a jet velocity of 0.26$c$, a jet kinetic power of 10$^{39}$ erg s$^{-1}$, a jet radius of 10 pc, and a molecular cloud density of 10$^{3}$ cm$^{-3}$, the shock velocity propagating within the cloud during a shock--cloud interaction is estimated to be $v_{\rm sh}$ $\sim$ 3 km s$^{-1}$. For a typical molecular clump size of $R_{\rm cl}$ $\sim$ 2 pc, the timescale required for the shock to traverse and disrupt the clump is given by $t_{\rm c}$ $\sim$ $R_{\rm cl}/v_{\rm sh}$ $\sim$ 0.7 Myr. Since this is significantly longer than the system age ($\sim$ 10$^{4}$--10$^{5}$ yr), the survival of molecular clumps in the jet head regions is not unexpected.

\cite{kayama2025} interpret the spectral properties of the X-ray jets based on the scenario proposed by \cite{sudoh2020}. In this framework, cosmic-ray electrons are accelerated in the head regions of the X-ray jets and subsequently lose energy via synchrotron cooling as they propagate downstream. In addition, localized regions with enhanced magnetic fields---such as the lenticular feature in the eastern jet and the w2 region in the western jet---are considered to exhibit enhanced synchrotron emissivity, where re-acceleration of electrons may occur. They constructed jet models that reproduce the observed spectra by adopting an electron injection spectral index of 2.08, a cutoff energy of the injection spectrum of 1.5 PeV, and jet velocities of 0.26$c$, 0.1$c$, and 0.065$c$ (i.e., one quarter of 0.26$c$; see also \citealt{kimura2020}). In these models, magnetic field strengths of 16 (9), 9 (7), and 7 (5) $\mu$G for the eastern (western) jet provide good agreement with the observed spectra in the head regions. Motivated by these results, we examine whether a shock--cloud interaction scenario can account for the observed extent of the X-ray jets dominated by synchrotron emission. 
As discussed in \cite{inoue2012}, shock--cloud interactions are capable of amplifying magnetic fields up to the mG level depending on the conditions. However, such strongly amplified fields are expected to be highly localized, and the synchrotron cooling timescales in these regions are correspondingly short (e.g., \citealt{uchiyama2007}). This implies that the average magnetic field strength is likely to be significantly lower. According to \cite{kimura2020}, using Equation (17) and assuming a jet kinetic power of $\sim$ 10$^{39}$ erg s$^{-1}$, the magnetic field strength is estimated to be $\sim$ 15 $\mu$G, which is comparable to the values adopted in \cite{kayama2025}.
Assuming that the magnetic fields in the head regions are amplified to the values derived by \cite{kayama2025}, and adopting the corresponding jet velocities, we estimate the propagation distance of cosmic-ray electrons responsible for the 2--7 keV emission observed with XMM-Newton, taking synchrotron cooling into account. We find that, in all cases considered, electrons emitting at 7 keV can propagate over distances comparable to or exceeding the size of the eastern and western X-ray jets ($\sim$ 41 pc) on synchrotron cooling timescales of $\sim$ 10$^{3}$ yr. In some cases, the estimated propagation distances exceed the observed extent of the X-ray jets. This discrepancy can be naturally explained by additional dissipation processes beyond simple synchrotron cooling, such as particle escape or enhanced synchrotron losses in localized regions of strong magnetic fields (e.g., the lenticular feature and the w1.5 and w2 regions). Therefore, the shock--cloud interaction scenario can naturally explain the large-scale structure of the X-ray jets dominated by synchrotron emission.

As noted above, very-high-energy (VHE) gamma rays have been detected from the X-ray jets of SS 433 by H.E.S.S. and LHAASO \citep{hess2024,lhaaso2025}. These emissions are likely of leptonic origin. The extent to which local magnetic field amplification in the head regions driven by shock--cloud interactions contributes to the acceleration of cosmic-ray electrons responsible for the VHE gamma-ray emission is not yet clear and is beyond the scope of this paper. Numerical simulations that incorporate turbulence driven by jet--cloud interactions, together with particle acceleration processes such as DSA, will be required to fully understand the origin of the high-energy emission.

On the observational side, high-resolution molecular-line studies with ALMA will be crucial for resolving the morphology and spatial extent of the molecular clumps. Observations of multiple molecular transitions, including optically thinner tracers, will also help constrain the physical conditions of the clumps and clarify the relationship between the CO-bright cores and the X-ray-emitting interaction layers.

\section{Conclusion} \label{sec:conclusion}
We report the first identification of molecular clumps likely associated with the re-brightening regions of the X-ray jets of SS 433, based on observations with the Nobeyama 45-m Radio Telescope. These clumps exhibit a clear spatial correlation with the X-ray emission, and the hardness ratio is enhanced at their surfaces.
These results indicate that the re-brightening of the X-ray jets is likely caused by magnetic field amplification due to jet-cloud interactions, which leads to enhanced synchrotron emission. 
Future high-resolution molecular-line observations will be essential for resolving the detailed morphology and physical conditions of the clumps and for further testing their physical association with the X-ray jets.
\begin{acknowledgments}
We are grateful to Dr. Naomi Tsuji for providing the Chandra datasets. 
We thank the anonymous referee for useful comments and constructive suggestions.
This study was supported by JSPS KAKENHI grant Nos. Haruka Sakemi: 22K20386, 23K13148, 26K17195, Hidetoshi Sano: 24H00246, Yasuo Fukui: 21H00040, 22H00152, Mami Machida: 22H01272, 23K22543, 24K00672, Shigeo S.Kimura: 23H04899, 26K00733, 26K00696, Masato I.N. Kobayashi: 22K14080, Kengo Tachihara: 20H01945, 23K20238. Shigeo S.Kimura acknowledges the support by the Tohoku Initiative for Fostering Global Researchers for Interdisciplinary Sciences (TI-FRIS) of MEXT's Strategic Professional Development Program for Young Researchers.
The Nobeyama 45-m radio telescope is operated by Nobeyama Radio Observatory, a branch of National Astronomical Observatory of Japan.
\end{acknowledgments}

\bibliography{sample701}{}
\bibliographystyle{aasjournalv7}



\end{document}